# Searching a systematics for nonfactorizable contribution to $B^-$ and $\bar{B}^0$ hadronic decays

Maninder Kaur[a], Supreet Pal Singh[b] and R. C. Verma[c]
*Department of Physics, Punjabi University,
Patiala – 147002, India.*

Abstract

Two-body weak decays $\bar{B} \to \pi D / \bar{B} \to \rho D$ and $\bar{B} \to \pi D^*$ are examined using isospin analysis to study nonfactorizable contributions. After determining the strong phases and obtaining the factorizable contributions from spectator-quark diagrams for $N_c=3$, we determine nonfactorizable isospin amplitudes from the experimental data for these modes. Our results support the universality of ratio of nonfactorizable isospin reduced amplitudes for these decays within experimental errors. In order to show that this systematics is not coincidental, we also plot our results w. r. t. this ratio.

[a] *maninderphy@gmail.com*
[b] *spsmudahar@gmail.com*
[c] *rcverma@gmail.com*

# I. INTRODUCTION

Experimental measurements for the weak decays of charm and bottom mesons have inspired several theoretical works to explore their dynamics [1-8]. The phenomenological analyses of two-body hadronic decays of heavy flavor mesons have indicated the presence of significant nonfactorizable contributions. In the naïve factorization scheme, the two QCD related coefficients $a_1$ and $a_2$ are treated as parameters to be fixed from the experimental data, while ignoring the nonfactorizable contribution to the decay amplitudes [9-11]. Initially, data on branching fractions of $D \to \bar{K}\pi$ decays seemed to require, $a_1 \approx c_1 = 1.26$, $a_2 \approx c_2 = -0.51$, leading to destructive interference between color-favored (CF) and color-suppressed (CS) processes for $D^+ \to \bar{K}^0 \pi^+$, thereby implying $N_c \to \infty$ limit [12-13]. This limit, which was thought to be justified with the hope that the nonfactorizable part relative to the factorizable amplitude is of the order of $1/N_c$, was expected to perform even better for the *B*- meson decays, where the final state particles carry larger momenta than that of the charm meson decays.

However, later the measurement of $\bar{B} \to D\pi$ meson decays did not favor this result empirically, as these decays require $a_1 \approx 1.03$, and $a_2 \approx 0.23$, *i.e.* a positive value of $a_2$, in sharp contrast to the expectations based on the large $N_c$ limit. Thus *B*- meson decays, revealing constructive interference between CF and CS diagrams for $B^- \to \pi^- D^0$, seem to favor $N_c = 3$ (real value). Even in the *D*-meson sector, the choice of the universal parameters $a_1$ and $a_2$ proved to be problematic when more accurate measurements were obtained for other decay modes of *D*-mesons even after including Final State Interactions (FSI) effects [14-15]. Consequently, the charm meson decays have been thoroughly reinvestigated to study the nonfactorization contributions explicitly. Using the isospin analysis for $D \to \bar{K}\pi / \bar{K}\rho / \bar{K}^*\pi$ decay modes [16-19], these contributions are expressed in terms of two reduced matrix elements $A_{1/2}^{nf}$, $A_{3/2}^{nf}$, and a systematics was recognized. It was observed that in all these decays, the nonfactorizable isospin reduced amplitude $A_{1/2}^{nf}$ not only has the same sign but also bear the same ratio (-1.12) respectively with $A_{3/2}^{nf}$ reduced amplitude, within the experimental errors. It is worth pointing out that this systematics was also found to be consistent with *p*-wave meson emitting decays of charm mesons: $D \to \bar{K}a_1 / \pi\bar{K}_1 / \pi\underline{\bar{K}}_1 / \pi\bar{K}_0 / \bar{K}a_2$ [18].

A lot of work has also been done to study the nonfactorization contributions in the charmed hadronic *B*-decays during the past two decades. The nonfactorizable terms may appear for several reasons, like FSI rescattering effects and soft-gluons exchange around the basic weak vertex. The rescattering effects on the outgoing mesons have been studied in detail for bottom meson decays [20-21]. Besides that, flavor *SU(3)* symmetry and Factorization Assisted Topological (FAT) approach have been employed for the study of such nonfactorizable contributions, as they have the advantage of absorbing various kinds of contributions lump- sum in terms of a few parameters, to be fixed empirically [22-23].

Inspired by these efforts, we investigate nonfactorizable contributions to the weak hadronic decays of bottom mesons, including the strong phases possibly through FSI. In fact it is well known that the strong phases of the decay amplitude in B-decays are quite significant, and lot of recent analysis of $\bar{B} \to \pi D / \rho D / \pi D^*$ decays have shown large strong phases. We perform the isospin analysis to study $\bar{B} \to \pi D / \rho D / \pi D^*$ decay modes. Aim of the present work is to investigate if such a systematics, which was observed in the charm meson decays, is valid for these decay modes, as their decay products, also have two different isospin states $I = 1/2$ and $3/2$. We include the strong phases which affect the interferences between isospin-1/2 and isospin-3/2 amplitudes. Using the experimental measurements for their branching fractions, we first obtain three free parameters the two isospin amplitudes $A_{1/2}$, $A_{3/2}$ and their relative strong phase, neglecting the overall strong phase. Determing the factorizable decay amplitudes for $N_c = 3$, we estimate the nonfactorizable isospin reduced amplitudes corresponding to these isospin states. We finally observe that the ratio of the nonfactorizable reduced amplitude in these isospin channels also follow a universal value for all the decay modes $\bar{B} \to \pi D / \bar{B} \to \rho D$ and $\bar{B} \to \pi D^*$.

## II. WEAK HAMILTONIAN

The effective weak Hamiltonian for CKM enhanced $B$-mesons decays is given by

$$H_w = \frac{G_F}{\sqrt{2}} V_{cb} V_{ud}^* \left[ c_1 \left( \bar{d}u \right) \left( \bar{c}b \right) + c_2 \left( \bar{c}u \right) \left( \bar{d}b \right) \right], \tag{1}$$

where $\bar{q}_1 q_2 = \bar{q}_1 \gamma_\mu (1 - \gamma_5) q_2$ denotes color singlet V−A Dirac current and the QCD coefficients [23-24] at bottom mass scale are

$$c_1 = 1.132, \qquad c_2 = -0.287. \tag{2}$$

Since the current operators in the weak Hamiltonian are expressed in terms of the fundamental quark fields, it is appropriate to have the Hamiltonian in a form such that one of these currents carries the same quantum numbers as one of the mesons emitted in the final state of bottom meson decays. Consequently the hadronic matrix elements of an operator O receives contributions from the operator itself and from the Fierz transformation of O, which generates the factorizable and nonfactorizable parts through the Fierz identity,

$$(\bar{d}u)(\bar{c}b) = \frac{1}{N_c} (\bar{c}u)(\bar{d}b) + \frac{1}{2} \left( \bar{c} \lambda^a u \right) \left( \bar{d} \lambda^a b \right), \tag{3}$$

where $\bar{q}_1 \lambda^a q_2 \equiv \bar{q}_1 \gamma_\mu (1 - \gamma_5) \lambda^a q_2$ represents the color octet current. Performing a similar treatment on the other operator $(\bar{c}u)(\bar{d}b)$, the weak Hamiltonian becomes

$$H_w^{CF} = \frac{G_F}{\sqrt{2}} V_{cb} V_{ud}^* \left[ a_1 \left( \bar{d}u \right)_H \left( \bar{c}b \right)_H + c_2 H_w^8 \right], \tag{4}$$

$$H_w^{CS} = \frac{G_F}{\sqrt{2}} V_{cb} V_{ud}^* \left[ a_2 \left( \bar{c}u \right)_H \left( \bar{d}b \right)_H + c_1 \tilde{H}_w^8 \right], \tag{5}$$

$$a_{1,2} = c_{1,2} + \frac{c_{2,1}}{N_c}, \tag{6}$$

$$H_w^8 = \frac{1}{2}\sum_{a=1}^{8}(\bar{c}\lambda^a u)(\bar{d}\lambda^a b), \qquad \tilde{H}_w^8 = \frac{1}{2}\sum_{a=1}^{8}(\bar{d}\lambda^a u)(\bar{c}\lambda^a b), \qquad (7)$$

to describe color-favored (CF) and color-suppressed (CS) processes, respectively. Here the index H in (4) and (5) indicates the change from quark current to hadron field operator [4]. The matrix elements of the first terms in (4) and (5) lead to the factorizable contributions [4] and the second terms, involving the color octet currents, generate nonfactorized contributions.

### III. DECAY MODES

#### A. $\bar{B} \to \pi D$ Decay mode

Branching fraction for *B*-meson decay into two pseudoscalar mesons is related to it decay amplitude, as

$$B(\bar{B} \to P_1 P_2) = \tau_B \left|\frac{G_F}{\sqrt{2}} V_{cb} V_{ud}^*\right|^2 \frac{p}{8\pi m_B^2}\left|A(\bar{B} \to P_1 P_2)\right|^2, \qquad (8)$$

where $\tau_B$ denotes the life time of *B*-mesons taken from [1],

$$\tau_{\bar{B}^0} = (1.519 \pm 0.004) \times 10^{-12} \ s, \qquad \tau_{B^-} = (1.638 \pm 0.004) \times 10^{-12} \ s,$$

$V_{ud}V_{cb}$ is the product of the Cabibbo–Kobayashi–Maskawa (CKM) matrix elements [1],

$$V_{ud} = 0.975, \qquad V_{cb} = 0.041,$$

and *p* is the magnitude of the 3-momentum of the final state particles in the rest frame of parent *B*- meson,

$$p = |p_1| = |p_2| = \frac{1}{2m_B}\left[\left\{m_B^2 - (m_1+m_2)^2\right\}\left\{m_B^2 - (m_1-m_2)^2\right\}\right]^{1/2}.$$

In heavy flavor meson decays, it has been observed that long distance FSI rescattering [20-21] of out-going mesons seriously affect their branching fractions. In general, such FSI phenomena can affect a decay amplitude in two ways. The decay amplitude may itself be modulated or it may acquire a phase. It has been shown by Kamal [25] that, in the weak scattering limit, elastic FSI– effect is mainly to pick up a phase factor, *i.e.,*

$$A^{FSI} = A\,e^{i\delta}. \qquad (9)$$

Consequently, mixing of final states, having the same quantum numbers, can take place. Initially, it was expected that bottom meson decays may not be affected by FSI as the produced particles may not have sufficient time to interact, and there are no meson resonances lying near *B*- meson mass corresponding to the quantum numbers of the final state. However, experimental data do not fulfill this naïve expectation [26].

To demonstrate this, we employ isospin framework in which $\bar{B} \to \pi D$ decay amplitudes are represented in terms of isospin reduced amplitudes including the strong interaction phases $\delta_{1/2}^{\pi D}$, $\delta_{3/2}^{\pi D}$ in respective Isospin -1/2 and 3/2 final states, as

$$A(\bar{B}^0 \to \pi^- D^+) = \frac{1}{\sqrt{3}}\left[ A_{3/2}^{\pi D} e^{i\delta_{3/2}^{\pi D}} + \sqrt{2} A_{1/2}^{\pi D} e^{i\delta_{1/2}^{\pi D}} \right],$$

$$A(\bar{B}^0 \to \pi^0 D^0) = \frac{1}{\sqrt{3}}\left[ \sqrt{2} A_{3/2}^{\pi D} e^{i\delta_{3/2}^{\pi D}} - A_{1/2}^{\pi D} e^{i\delta_{1/2}^{\pi D}} \right], \qquad (10)$$

$$A(B^- \to \pi^- D^0) = \sqrt{3} A_{3/2}^{\pi D} e^{i\delta_{3/2}^{\pi D}}.$$

These lead to the following relations:

$$A_{1/2}^{\pi D} = \left[ \left|A(\bar{B}^0 \to \pi^- D^+)\right|^2 + \left|A(\bar{B}^0 \to \pi^0 D^0)\right|^2 - \frac{1}{3}\left|A(B^- \to \pi^- D^0)\right|^2 \right]^{1/2},$$

$$A_{3/2}^{\pi D} = \sqrt{\frac{1}{3}} \left|A(B^- \to \pi^- D^0)\right|, \qquad (11)$$

and the relative phase difference, $\delta^{\pi D} = \delta_{1/2}^{\pi D} - \delta_{3/2}^{\pi D}$, is given by

$$\cos\delta^{\pi D} = \frac{(3\left|A(\bar{B}^0 \to \pi^- D^+)\right|^2 - 6\left|A(\bar{B}^0 \to \pi^0 D^0)\right|^2 + \left|A(B^- \to \pi^- D^0)\right|^2)}{6\sqrt{2}\left|A_{1/2}^{\pi D}\right|\left|A_{3/2}^{\pi D}\right|}. \qquad (12)$$

Thus $A_{1/2}^{\pi D}$ and $A_{3/2}^{\pi D}$ can be treated as real quantities in the following analysis.

Using the experimental values [1],

$$B(\bar{B}^0 \to \pi^- D^+) = (2.52 \pm 0.13) \times 10^{-3},$$
$$B(\bar{B}^0 \to \pi^0 D^0) = (2.63 \pm 0.14) \times 10^{-4},$$
$$B(B^- \to \pi^- D^0) = (4.68 \pm 0.13) \times 10^{-3},$$

we thus obtain

$$A_{1/2}^{\pi D\ \exp} = \pm(1.273 \pm 0.065)\,GeV^3, \qquad A_{3/2}^{\pi D\ \exp} = \pm(1.323 \pm 0.018)\,GeV^3, \qquad (13)$$

and the phase difference

$$\delta^{\pi D} = (28 \pm 7)^\circ, \qquad (14)$$

which agrees with the final state rescattering analysis [22]. Though this phase difference is relative smaller than that of $D \to \bar{K}\pi$ mode $\delta = (86 \pm 7)^\circ$, it certainly shows the presence of non-vanishing strong phases in the *B*- meson sector.

We express total decay amplitude as sum of the factorizable and nonfactorizable parts,

$$A(\bar{B} \to \pi D) = A^f(\bar{B} \to \pi D) + A^{nf}(\bar{B} \to \pi D), \quad (15)$$

arising from the respective terms of the weak Hamiltonian given in (4) and (5).

Using the factorization scheme, spectator- quark parts of the decay amplitudes arising from $W$- emission[†] diagrams are derived for the following classes of $\bar{B} \to \pi D$ decays:

a. Class I: Color favored (CF)

$$A^f(\bar{B}^0 \to \pi^- D^+) = a_1 f_\pi (m_B^2 - m_D^2) F_0^{\bar{B}D}(m_\pi^2), \quad (16)$$

b. Class II: Color Suppressed (CS)

$$A^f(\bar{B}^0 \to \pi^0 D^0) = -\frac{1}{\sqrt{2}} a_2 f_D (m_B^2 - m_\pi^2) F_0^{\bar{B}\pi}(m_D^2), \quad (17)$$

c. Class III : Interference of CF and CS

$$A^f(B^- \to \pi^- D^0) = a_1 f_\pi (m_B^2 - m_D^2) F_0^{\bar{B}D}(m_\pi^2) + a_2 f_D (m_B^2 - m_\pi^2) F_0^{\bar{B}\pi}(m_D^2), \quad (18)$$

where $a_1 = c_1 + \dfrac{c_2}{N_c}$, and $a_2 = c_2 + \dfrac{c_1}{N_c}$.

We calculate values of the factorization contributions for $N_c = 3$ (real value) using numerical inputs for decay constants taken as,

$$f_D = (0.207 \pm 0.009) GeV, \; f_\pi = (0.131 \pm 0.002) GeV, \quad (19)$$

from the leptonic decays of $D$ and $\pi$ mesons, respectively [27].

---

[†] In general, $W$- exchange, $W$- annihilation and $W$- loop diagrams may also contribute to the bottom meson decays. Note that $W$-annihilation and $W$-loop processes do not appear for any of the decays considered in this work. $W$- exchange is usually suppressed due to helicity and color arguments, for which the partial decay rate depends on the wave function at the origin, and in the relative ratio of its contribution to that of the spectator diagrams is given by

$$\frac{\Gamma_{W-exc/anni}}{\Gamma_{spect}} \approx \frac{|\psi(0)|^2}{m^3} \approx \alpha_s^3 \left(\frac{m_q}{M_Q}\right)^3,$$

where $m_q$ and $M_Q$ represents masses of the light and heavy quark in the $B$-mesons, As the mass of heavy quark goes up, these become less and less important [4]. Particularly for $\bar{B} \to D\pi$ decays, it has been categorically shown by Kamal and Pham [11] that $W$- exchange terms are highly suppressed due to smallness of the relevant form-factor $F_0^{D\pi}(m_B^2)$, and of the color factor $a_2$. Recently, this observation has further been supported in the FAT based analysis of these decays [23], so contribution of $W$- exchange diagram can be neglected specially in the presence of $W$-emission diagram.

Assuming the nearest pole dominance, momentum dependence of the form-factors, appearing in the decay amplitudes given in (17-19), is taken as

$$F_0(q^2) = \frac{F_0(0)}{\left(1 - q^2/m_s^2\right)^n},$$
(20)

where pole masses are given by the scalar meson carrying the quantum number of the corresponding weak current, which are $m_s = 5.78$ GeV, and $m_s = 6.80$ GeV, and $n = 1$ for the monopole formula. Form-factors $F_0(0)$ at $q^2 = 0$ are taken from [28], as given below,

$$F_0^{\bar{B}\pi}(0) = (0.27 \pm 0.05),$$
$$F_0^{\bar{B}D}(0) = (0.66 \pm 0.03).$$
(21)

We finally get,

$$A^f(\bar{B}^0 \to \pi^- D^+) = 2.180 \pm 0.099 \quad GeV^3,$$
$$A^f(\bar{B}^0 \to \pi^0 D^0) = -0.111 \pm 0.021 \quad GeV^3,$$
$$A^f(B^- \to \pi^- D^0) = 2.339 \pm 0.103 \quad GeV^3.$$
(22)

Exploiting the following isospin relations:

$$A_{1/2}^f(\bar{B} \to \pi D) = \frac{1}{\sqrt{3}} \left\{ \sqrt{2} A^f(\bar{B}^0 \to \pi^- D^+) - A^f(\bar{B}^0 \to \pi^0 D^0) \right\},$$
$$A_{3/2}^f(\bar{B} \to \pi D) = \frac{1}{\sqrt{3}} \left\{ A^f(\bar{B}^0 \to \pi^- D^+) + \sqrt{2} A^f(\bar{B}^0 \to \pi^0 D^0) \right\},$$
(23)

we obtain

$$A_{1/2}^f = (1.845 \pm 0.082) GeV^3, \qquad A_{3/2}^f = (1.168 \pm 0.060) GeV^3.$$
(24)

Using isospin C. G. coefficients with the convention used in [17, 18], nonfactorizable part of the decay amplitudes can be expressed in terms of the scattering amplitudes for spurion $+ \bar{B} \to \pi D$ process:

$$A^{nf}(\bar{B}^0 \to \pi^- D^+) = \frac{1}{3} c_2 \left( \langle \pi D \| H_w^8 \| \bar{B} \rangle_{3/2} + 2 \langle \pi D \| H_w^8 \| \bar{B} \rangle_{1/2} \right),$$
$$A^{nf}(\bar{B}^0 \to \pi^0 D^0) = \frac{\sqrt{2}}{3} c_1 \left( \langle \pi D \| \tilde{H}_w^8 \| \bar{B} \rangle_{3/2} - \langle \pi D \| \tilde{H}_w^8 \| \bar{B} \rangle_{1/2} \right),$$
$$A^{nf}(B^- \to \pi^- D^0) = c_2 \langle \pi D \| H_w^8 \| \bar{B} \rangle_{3/2} + c_1 \langle \pi D \| \tilde{H}_w^8 \| \bar{B} \rangle_{3/2}.$$
(25)

At present, there is no available technique to calculate these quantities exactly from the theory of strong interactions. Therefore, subtracting the factorizable part (24) from the experimental decay amplitude (13), we determine the nonfactorizable isospin reduced amplitudes,

$$A^{nf}_{1/2} = -(0.572 \pm 0.105)\,GeV^3, \quad A^{nf}_{3/2} = -(2.491 \pm 0.062)\,GeV^3, \tag{26}$$

by choosing positive and negative values for $A^{nf}_{1/2}$ and $A^{nf}_{3/2}$, respectively. Their ratio is given below

$$\alpha = \frac{A^{nf}_{1/2}}{A^{nf}_{3/2}} = 0.229 \pm 0.042. \tag{27}$$

There are several existing calculations for the form factors obtained from different approaches in the literature, which are given in the Table 1.

Table 1. Form-factor of $\bar{B} \to D$ and $\bar{B} \to \pi$ transitions at maximum recoil ($q^2 = 0$).

| Form - factor | CLFQM [27] | LQCD [28] | LCSR [29] | pQCD [30] |
|---|---|---|---|---|
| $F_0^{\bar{B}D}(0)$ | 0.67±0.01 | 0.66±0.03 | 0.65±0.08 | 0.673±0.063 |
| $F_0^{\bar{B}\pi}(0)$ | 0.25±0.01 | 0.27±0.05 | 0.21±0.07 | --- |

In order to see the effect of variation of the form-factors on our analysis, we give ratio $\alpha$ in Table 2 for maximum and minimum values of the form-factors, which are consistent with (27) with in errors.

Table 2. Ratio $\alpha = \dfrac{A^{nf}_{1/2}}{A^{nf}_{3/2}}$ for maximum and minimum values of form-factors.

| $F_0^{\bar{B}D}(0)$ | 0.67 | 0.67 | 0.65 | 0.65 |
|---|---|---|---|---|
| $F_0^{\bar{B}\pi}(0)$ | 0.27 | 0.21 | 0.27 | 0.21 |
| $\alpha$ | 0.238 | 0.231 | 0.220 | 0.213 |

We also presents the dependence of $\alpha$ on form-factor $F_0^{\bar{B}D}(0)$ and $F_0^{\bar{B}\pi}(0)$ in Figure 1.

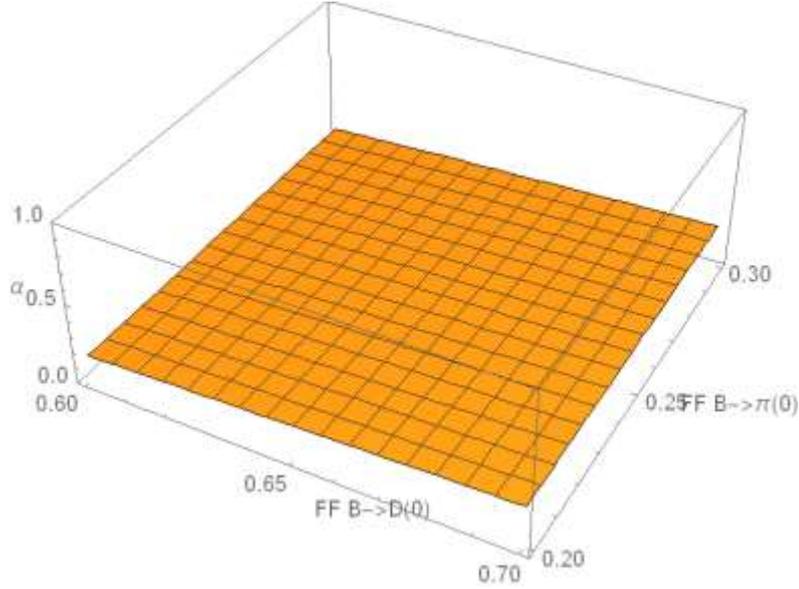

Figure 1. Variation of $\alpha$ with form factors $F_0^{\bar{B}D}(0)$ and $F_0^{\bar{B}\pi}(0)$.

B. $\bar{B} \to \rho D$ Decay mode

Using the branching fraction,

$$B(\bar{B} \to PV) = \tau_B \left| \frac{G_F}{\sqrt{2}} V_{cb} V_{ud}^* \right|^2 \frac{p^3}{8\pi m_V^2} \left| A(\bar{B} \to PV) \right|^2, \qquad (28)$$

where,

$$p = |p_1| = |p_2| = \frac{1}{2m_B} \left\{ \left( m_B^2 - (m_P + m_V)^2 \right) \left( m_B^2 - (m_P - m_V)^2 \right) \right\}^{1/2}.$$

Since isospin structure of $\bar{B} \to \rho D$ decays is exactly the same as that of $\bar{B} \to \pi D$ decays, amplitudes are represented in terms of isospin reduced amplitudes including the strong interaction phases $\delta_{1/2}^{\rho D}$, $\delta_{3/2}^{\rho D}$ in respective Isospin -1/2 and 3/2 final states, as

$$\begin{aligned} A(\bar{B}^0 \to \rho^- D^+) &= \frac{1}{\sqrt{3}} \left[ A_{3/2}^{\rho D} e^{i\delta_{3/2}^{\rho D}} + \sqrt{2} A_{1/2}^{\rho D} e^{i\delta_{1/2}^{\rho D}} \right], \\ A(\bar{B}^0 \to \rho^0 D^0) &= \frac{1}{\sqrt{3}} \left[ \sqrt{2} A_{3/2}^{\rho D} e^{i\delta_{3/2}^{\rho D}} - A_{1/2}^{\rho D} e^{i\delta_{1/2}^{\rho D}} \right], \qquad (29) \\ A(B^- \to \rho^- D^0) &= \sqrt{3} A_{3/2}^{\rho D} e^{i\delta_{1/2}^{\rho D}}, \end{aligned}$$

yielding

$$A_{1/2}^{\rho D} = \left[ \left|A(\bar{B}^0 \to \rho^- D^+)\right|^2 + \left|A(\bar{B}^0 \to \rho^0 D^0)\right|^2 - \frac{1}{3}\left|A(B^- \to \rho^- D^0)\right|^2 \right]^{1/2},$$

$$A_{3/2}^{\rho D} = \sqrt{\frac{1}{3}}\left|A(B^- \to \rho^- D^0)\right|,$$

(30)

and the relative phase difference, $\delta^{\rho D} = \delta_{1/2}^{\rho D} - \delta_{3/2}^{\rho D}$, given by

$$\cos\delta^{\rho D} = \frac{\left(3\left|A(\bar{B}^0 \to \rho^- D^+)\right|^2 - 6\left|A(\bar{B}^0 \to \rho^0 D^0)\right|^2 + \left|A(B^- \to \rho^- D^0)\right|^2\right)}{6\sqrt{2}\left|A_{1/2}^{\rho D}\right|\left|A_{3/2}^{\rho D}\right|}.$$

(31)

From the experimental branching fractions,

$$B(\bar{B}^0 \to \rho^- D^+) = (7.6 \pm 1.2) \times 10^{-3},$$
$$B(\bar{B}^0 \to \rho^0 D^0) = (3.21 \pm 0.21) \times 10^{-4},$$
$$B(B^- \to \rho^- D^0) = (1.34 \pm 0.18) \times 10^{-2},$$

we calculate the total isospin reduced amplitudes:

$$A_{1/2}^{\rho D \, \text{exp}} = \pm(0.143 \pm 0.025) \, GeV^2, \qquad A_{3/2}^{\rho D \, \text{exp}} = \pm(0.149 \pm 0.010) \, GeV^2,$$

(32)

and the phase difference between I=1/2 and 3/2 final states,

$$\delta^{\rho D} = \left(8 \, {}^{+30}_{-8}\right)^\circ.$$

(33)

Factorizable decay amplitudes for the spectator- quark diagrams can be written as:

$$A^f(\bar{B}^0 \to \rho^- D^+) = 2a_1 m_\rho f_\rho F_1^{\bar{B}D}(m_\rho^2),$$
$$A^f(\bar{B}^0 \to \rho^0 D^0) = -\sqrt{2} a_2 f_D m_\rho A_0^{\bar{B}\rho}(m_D^2),$$
$$A^f(B^- \to \rho^- D^0) = a_1 2m_\rho f_\rho F_1^{\bar{B}D}(m_\rho^2) + a_2 f_D 2m_\rho A_0^{\bar{B}\rho}(m_D^2).$$

(34)

It has been pointed out in the BSW II model [3] that consistency with the heavy quark symmetry requires certain form- factors, such as $F_1(0)$ and $A_0(0)$, to have dipole $q^2$ dependence (*n=2*) in,

$$F_1(q^2) = \frac{F_1(0)}{\left(1 - {q^2}/{m_V^2}\right)^n}, \qquad A_0(q^2) = \frac{A_0(0)}{\left(1 - {q^2}/{m_P^2}\right)^n},$$

(35)

where vector $V(1^-)$ meson and pseudoscalar $P(0^-)$ meson pole masses are 6.34 *GeV* and 5.27 *GeV*, respectively.

Decay constants values are taken from [27],

$$f_D = (0.207 \pm 0.009)\,GeV, \quad f_\rho = (0.215 \pm 0.005)\,GeV, \tag{36}$$

and form-factors for $\bar{B} \to V$ transitions are chosen from [31],

$$A_0^{\bar{B}\rho}(0) = 0.356 \pm 0.042, \tag{37}$$

where $F_0^{\bar{B}D}(0)$ value is taken from equation (21)

$$F_1^{\bar{B}D}(0) = F_0^{\bar{B}D}(0) = 0.66 \pm 0.03. \tag{38}$$

Thus, we calculate the factorizable contributions to the decay amplitudes,

$$\begin{aligned}
A^f(\bar{B}^0 \to \rho^- D^+) &= (0.235 \pm 0.011)\,GeV^2, \\
A^f(\bar{B}^0 \to \rho^0 D^0) &= -(0.010 \pm 0.001)\,GeV^2, \\
A^f(B^- \to \rho^- D^0) &= (0.248 \pm 0.011)\,GeV^2,
\end{aligned} \tag{39}$$

thereby isospin reduced amplitudes of factorized amplitudes are calculated as,

$$A_{1/2}^f = (0.197 \pm 0.009)\,GeV^2, \qquad A_{3/2}^f = (0.127 \pm 0.006)\,GeV^2. \tag{40}$$

Following the procedure discussed for $\bar{B} \to \pi D$, we subtract the factorizable parts (40) from the experimental decay amplitudes (32) to determine nonfactorizable reduced isospin amplitudes,

$$A_{1/2}^{nf} = -(0.054 \pm 0.026)\,GeV^2, \qquad A_{3/2}^{nf} = -(0.277 \pm 0.012)\,GeV^2, \tag{41}$$

which bear the following ratio:

$$\alpha = \frac{A_{1/2}^{nf}}{A_{3/2}^{nf}} = 0.200 \pm 0.096. \tag{42}$$

There are some existing calculations for $A_0^{\bar{B}\rho}(0)$ also which are given in Table 3. To show the effect of form-factor on our analysis we obtain ratio $\alpha = \dfrac{A_{1/2}^{nf}}{A_{3/2}^{nf}}$ for maximum and minimum value of the form factors given in Table 4, which are consistent with (42) within errors.

Table 3. Form-factor of the $\bar{B} \to \rho$ transitions at maximum recoil ($q^2$ =0).

| Form-factor | CLFQM [27] | LCSR [31] | LCSR [32] | CLFQM [33] | PQCD [34] |
|---|---|---|---|---|---|
| $A_0^{\bar{B}\rho}(0)$ | 0.32±0.01 | 0.356±0.042 | 0.303 | 0.30±0.05 | 0.366±0.036 |

Table 4. Ratio $\alpha = \dfrac{A_{1/2}^{nf}}{A_{3/2}^{nf}}$ for maximum and minimum values of form-factors.

| | | | | |
|---|---|---|---|---|
| $F_0^{\bar{B}D}(0)$ | 0.67 | 0.67 | 0.65 | 0.65 |
| $A_0^{\bar{B}\rho}(0)$ | 0.40 | 0.30 | 0.40 | 0.30 |
| $\alpha$ | 0.208 | 0.201 | 0.190 | 0.183 |

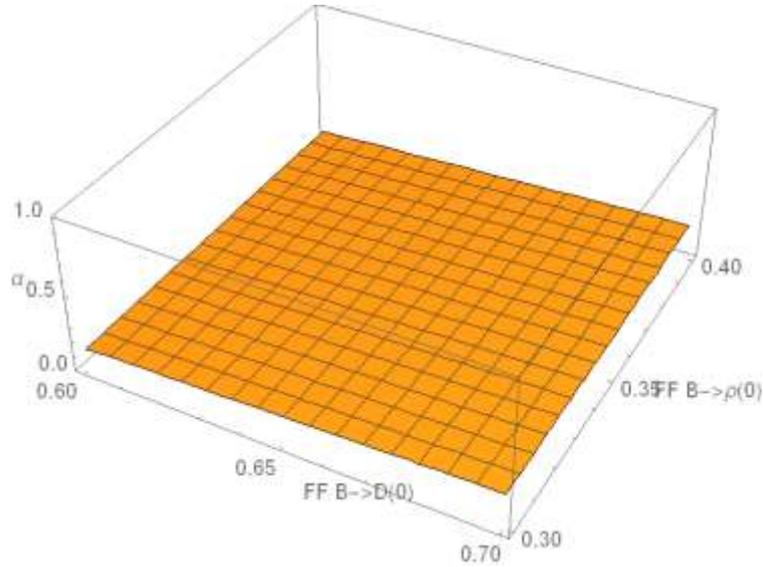

Figure 2. Variation of $\alpha$ with form factors $F_0^{\bar{B}D}(0)$ and $F_0^{\bar{B}\rho}(0)$.

### C. $\bar{B} \to \pi D^*$ Decay mode

Including the strong phases among for isospin I=1/2 and 3/2 states, the decay amplitudes are given by

$$A(\bar{B}^0 \to \pi^- D^{*+}) = \frac{1}{\sqrt{3}}\left[ A_{3/2}^{\pi D^*} e^{i\delta_{3/2}^{\pi D^*}} + \sqrt{2} A_{1/2}^{\pi D^*} e^{i\delta_{1/2}^{\pi D^*}} \right],$$

$$A(\bar{B}^0 \to \pi^0 D^{*0}) = \frac{1}{\sqrt{3}}\left[ \sqrt{2} A_{3/2}^{\pi D^*} e^{i\delta_{3/2}^{\pi D^*}} - A_{1/2}^{\pi D^*} e^{i\delta_{1/2}^{\pi D^*}} \right], \quad (43)$$

$$A(B^- \to \pi^- D^{*0}) = \sqrt{3} A_{3/2}^{\pi D^*} e^{i\delta_{3/2}^{\pi D^*}}.$$

These lead to the following reduced isospin amplitudes:

$$A_{1/2}^{\pi D^*} = \left[ \left|A(\bar{B}^0 \to \pi^- D^{*+})\right|^2 + \left|A(\bar{B}^0 \to \pi^0 D^{*0})\right|^2 - \frac{1}{3}\left|A(B^- \to \pi^- D^{*0})\right|^2 \right]^{1/2},$$

$$A_{3/2}^{\pi D^*} = \sqrt{\frac{1}{3}}\left|A(B^- \to \pi^- D^{*0})\right|, \quad (44)$$

and the relative phase difference, $\delta^{\pi D^*} = \delta_{1/2}^{\pi D^*} - \delta_{3/2}^{\pi D^*}$, given by

$$\cos\delta^{\pi D^*} = \frac{(3\left|A(\bar{B}^0 \to \pi^- D^{*+})\right|^2 - 6\left|A(\bar{B}^0 \to \pi^0 D^{*0})\right|^2 + \left|A(B^- \to \pi^- D^{*0})\right|^2)}{6\sqrt{2}\left|A_{1/2}^{\pi D^*}\right|\left|A_{3/2}^{\pi D^*}\right|}. \quad (45)$$

Using the experimental values of branching fractions [1],

$$B(\bar{B}^0 \to \pi^- D^{*+}) = (2.74 \pm 0.13) \times 10^{-3},$$
$$B(\bar{B}^0 \to \pi^0 D^{*0}) = (2.20 \pm 0.60) \times 10^{-4},$$
$$B(B^- \to \pi^- D^{*0}) = (4.90 \pm 0.17) \times 10^{-3},$$

we calculate total isospin reduced amplitudes

$$A_{1/2}^{\pi D^* \, \exp} = \pm(0.226 \pm 0.042)\, GeV^2, \quad A_{3/2}^{\pi D^* \, \exp} = \pm(0.231 \pm 0.040)\, GeV^2, \quad (46)$$

and the phase difference

$$\delta^{\pi D^*} = (24 \pm 24)^\circ. \quad (47)$$

So the $\bar{B} \to PV$ decays also indicate the presence of strong FSI phases as has also been observed in [17-18].

The factorizable amplitudes for this mode are:

$$A^f(\bar{B}^0 \to \pi^- D^{*+}) = 2a_1 m_{D^*} f_\pi A_0^{\bar{B}D^*}(m_\pi^2),$$

$$A^f(\bar{B}^0 \to \pi^0 D^{*0}) = -\sqrt{2} a_2 f_{D^*} m_{D^*} F_1^{\bar{B}\pi}(m_{D^*}^2), \quad (48)$$

$$A^f(B^- \to \pi^- D^{*0}) = a_1 2 m_{D^*} f_\pi A_0^{\bar{B}D^*}(m_\pi^2) + a_2 f_{D^*} 2 m_{D^*} F_1^{\bar{B}\pi}(m_{D^*}^2).$$

Using the decay constant values [27],

$$f_{D^*} = (0.245 \pm 0.034)\,GeV, \qquad f_\pi = (0.131 \pm 0.002)\,GeV, \qquad (49)$$

and the form-factor,

$$A_0^{\bar{B}D^*}(0) = (0.68 \pm 0.04), \qquad (50)$$

taken from [27], with pole masses $V(1^-) = 5.32$ $GeV$ and $P(0^-) = 6.28$ $GeV$ for $q^2$-dependence (35),

$$F_1^{\bar{B}\pi}(0) = F_0^{\bar{B}\pi}(0) = (0.27 \pm 0.05),$$

we calculate the factorized amplitudes as given below,

$$A^f(\bar{B}^0 \to \pi^- D^{*+}) = (0.371 \pm 0.022)\,GeV^2,$$
$$A^f(\bar{B}^0 \to \pi^0 D^{*0}) = -(0.023 \pm 0.004)\,GeV^2, \qquad (51)$$
$$A^f(B^- \to \pi^- D^{*0}) = (0.403 \pm 0.023)\,GeV^2,$$

which inturn yield the isospin reduced amplitudes,

$$A_{1/2}^f = (0.317 \pm 0.018)\,GeV^2, \qquad A_{3/2}^f = (0.196 \pm 0.013)\,GeV^2. \qquad (52)$$

Subtracting factorizable parts from the total experimental amplitudes, we calculate

$$A_{1/2}^{nf} = -(0.090 \pm 0.046)\,GeV^2, \qquad A_{3/2}^{nf} = -(0.426 \pm 0.042)\,GeV^2, \qquad (53)$$

having the following ratio:

$$\alpha = \frac{A_{1/2}^{nf}}{A_{3/2}^{nf}} = 0.211 \pm 0.109. \qquad (54)$$

In literature, we find different values of $A_0^{\bar{B}D^*}(0)$ form-factor as shown in Table 5, and calculate the ratio $\alpha$ for maximum and minimum values of the form-factors, and plot the variation of $\alpha$ in Figure 3.

Table 5. Form-factor of the $\bar{B} \to D^*$ transitions at maximum recoil ($q^2=0$).

| Form-factor | CLFQM [27] | CLFQM [33] | LQCD [35] |
|---|---|---|---|
| $A_0^{\bar{B}D^*}(0)$ | 0.68±0.04 | 0.68±0.08 | 0.921±0.013 |

Table 6. Ratio $\alpha = \dfrac{A_{1/2}^{nf}}{A_{3/2}^{nf}}$ for maximum and minimum values of form-factors.

| $A_0^{\bar{B}D^*}(0)$ | 0.92 | 0.92 | 0.68 | 0.68 |
|---|---|---|---|---|
| $F_0^{\bar{B}\pi}(0)$ | 0.27 | 0.21 | 0.27 | 0.21 |
| $\alpha$ | 0.392 | 0.382 | 0.211 | 0.203 |

This shows significant variation due to uncertainties in $A_0^{\bar{B}D^*}(0)$, we feel that higher value of $A_0^{\bar{B}D^*}(0)$ is less likely, which may be determined more precisely in future.
.

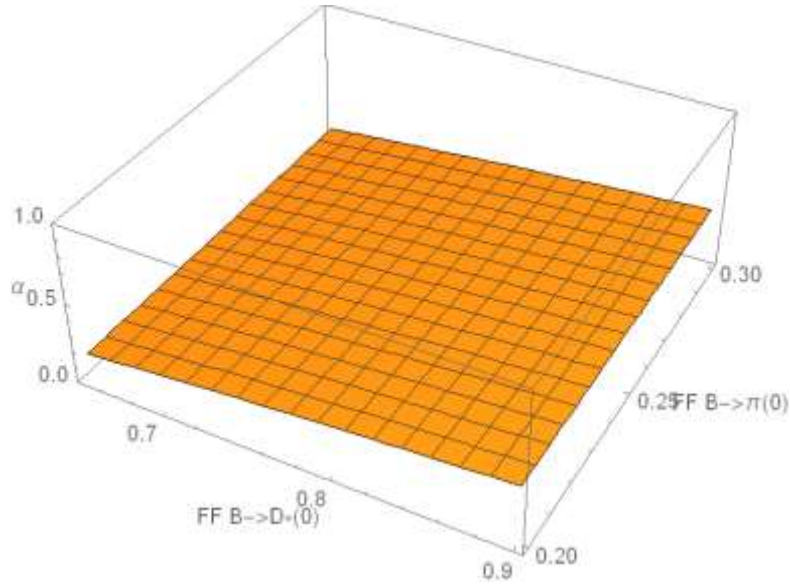

Figure 3. Variation of $\alpha$ with form factors $A_0^{\bar{B}D^*}(0)$ and $F_0^{\bar{B}\pi}(0)$

## IV. RESULTS AND DISCUSSIONS

The purpose of performing the isospin analysis of $\bar{B} \to \pi D$ and $\bar{B} \to \rho D / \pi D^*$ decays has been to search for the systematics, which has been recognized in the charm sector before [17-18]. We observe that by choosing positive sign of $A_{1/2}^{\text{exp}}$ and negative sign for $A_{3/2}^{\text{exp}}$ in each case, we get the same value for the ratio of corresponding nonfactorizable reduced matrix elements $A_{1/2}^{nf}$ and $A_{3/2}^{nf}$, *i.e.,*

$$\frac{A^{nf}_{1/2}(\bar{B}\to\pi D)}{A^{nf}_{3/2}(\bar{B}\to\pi D)} = \frac{A^{nf}_{1/2}(\bar{B}\to\rho D)}{A^{nf}_{3/2}(\bar{B}\to\rho D)} = \frac{A^{nf}_{1/2}(\bar{B}\to\pi D^*)}{A^{nf}_{3/2}(\bar{B}\to\pi D^*)},$$

$$0.229\pm 0.042 \quad 0.200\pm 0.096 \quad 0.211\pm 0.109$$

(55)

and, notice that $A^{nf}_{1/2}$ has negative sign for all the cases,

$$A^{nf}_{1/2}(\bar{B}\to\pi D) = -(0.572\pm 0.105)\,GeV^3, \tag{56}$$

$$A^{nf}_{1/2}(\bar{B}\to\rho D) = -(0.054\pm 0.026)\,GeV^2, \tag{57}$$

$$A^{nf}_{1/2}(\bar{B}\to\pi D^*) = -(0.090\pm 0.046)\,GeV^2. \tag{58}$$

We can predict sum of the branching fractions of the $\bar{B}^0$ – meson decays, in respective modes considered here, generically as,

$$B_{-+} + B_{00} = \frac{\tau_{\bar{B}^0}}{3\tau_{B^-}} B_{0-}\left[1 + \left\{\alpha + \frac{(\sqrt{2}-\alpha)A^f_{-+} - (1+\sqrt{2}\alpha)A^f_{00}}{A_{0-}}\right\}^2\right], \tag{59}$$

where $\alpha$ is defined already (27), and experimental decay amplitude of $B^-$ decays,

$$A_{0-} = \sqrt{\frac{B_{0-}}{\tau_{B^-}\times(kinematic\ factor)}},$$

(subscripts -+, 00, 0- denote the charge states of the non-charm and charm mesons emitted in each case and $\tau$ denotes the life time). Taking, the average value of $\alpha = 0.22$, we predict

$$B(\bar{B}^0\to\pi^- D^+) + B(\bar{B}^0\to\pi^0 D^0) = (0.28\pm 0.02)\%\quad Theo,$$
$$= (0.28\pm 0.01)\%\quad Expt; \tag{60}$$

$$B(\bar{B}^0\to\rho^- D^+) + B(\bar{B}^0\to\rho^0 D^0) = (0.76\pm 0.13)\%\quad Theo,$$
$$= (0.79\pm 0.12)\%\quad Expt; \tag{61}$$

$$B(\bar{B}^0\to\pi^- D^{*+}) + B(\bar{B}^0\to\pi^0 D^{*0}) = (0.29\pm 0.04)\%\quad Theo,$$
$$= (0.30\pm 0.01)\%\quad Expt; \tag{62}$$

which are in good agreement with the experiment. To show that this agreement is not coincidental and to study the sensitivity of the sum of the $\bar{B}^0$ branching fractions with ratio $\alpha$, we plot $\sum B(\bar{B}^0\to decays)$ against $\alpha$ by treating it as a free parameter for all the three cases which are shown in Figure 4, 5, and 6. Clearly, the experiment data indicate $\alpha = 0.22$ consistently.

We wish to remark that similar observations have also been made in the FAT approach [23] analysis used for *B*- meson decays, which separates the factorizable and nonfactorizable contributions in each topological quark level diagrams. The most important result in this approach is that the non-perturbative parameters, $\chi^{C,E}$ and $\varphi^{C,E}$ representing the nonfactorizable contributions, are found to be universal for all the $\bar{B} \to \pi D / \rho D / \pi D^*$ decay modes, which is consistent with systematics recognized in our analysis.

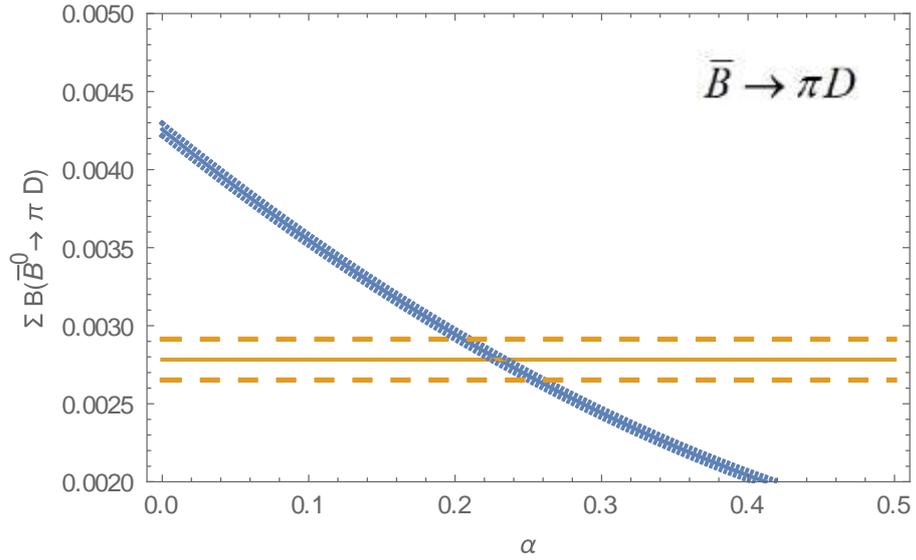

Figure 4. Variation of the sum of $B(\bar{B}^0 \to \pi^- D^+)$ and $B(\bar{B}^0 \to \pi^0 D^0)$ with the ratio $\alpha = A^{nf}_{1/2} / A^{nf}_{3/2}$. The broken curves show the errors due to decay constant, form factors and $B(B^- \to \pi^- D^0)$. The horizontal lines correspond to experimental value for the sum and its errors are shown by broken lines.

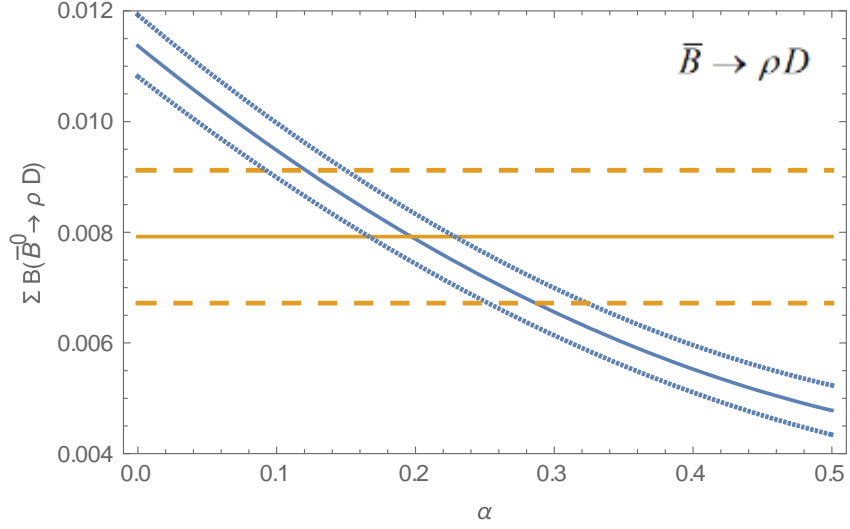

Figure 5. Variation of the sum of $B(\bar{B}^0 \to \rho^- D^+)$ and $B(\bar{B}^0 \to \rho^0 D^0)$ with the ratio $\alpha = A_{1/2}^{nf}/A_{3/2}^{nf}$. The broken curves show the errors due to decay constant, form factors and $B(B^- \to \rho^- D^0)$. The horizontal lines correspond to experimental value for the sum and its errors are shown by broken lines.

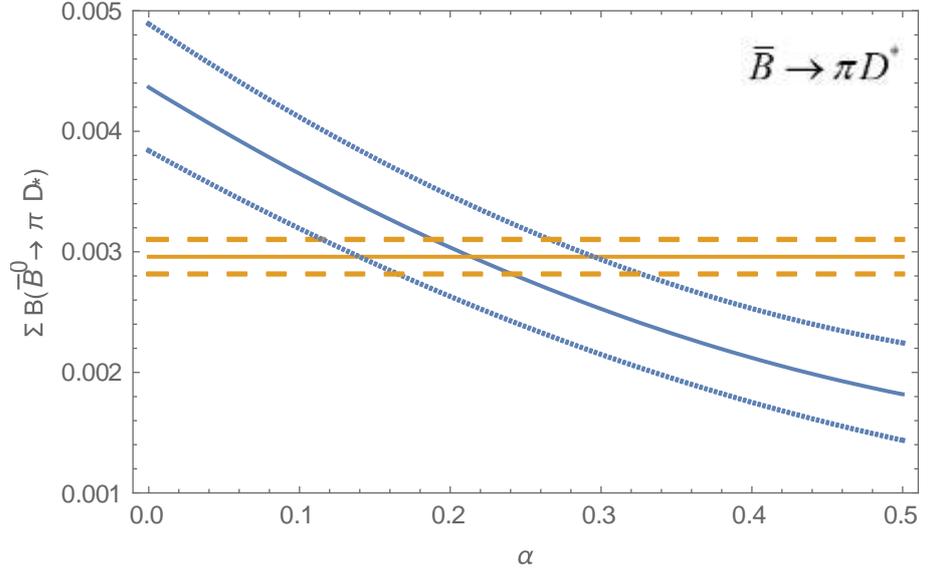

Figure 6. Variation of the sum of $B(\bar{B}^0 \to \pi^- D^{*+})$ and $B(\bar{B}^0 \to \pi^0 D^{*0})$ with the ratio $\alpha = A_{1/2}^{nf}/A_{3/2}^{nf}$. The broken curves show the errors due to due to decay constant, form factors and $B(B^- \to \pi^- D^{*0})$. The horizontal lines correspond to experimental value for the sum and its errors are shown by dotted lines.

## V. SUMMARY AND CONCLUSIONS

We have carried out an analysis of CKM-favored two-body hadronic decays $\bar{B} \to \pi D / \rho D / \pi D^*$, which involve two isospin states in the decay products, by including nonfactorizable contributions arising due to part of the weak Hamiltonian involving the colored currents. Since the nonfactorizable contributions being non-perturbative are difficult to be calculated. At present from the theory of strong interactions, we have employed the isospin formalism and identified that in all the decay modes, the nonfactorizable isospin reduced amplitude $A_{1/2}^{nf}$ bears the same ratio 0.21 with $A_{3/2}^{nf}$ consistently, with in the experimental errors, as well as keeping the same sign. It is significant to point out that similar universality in nonfactorizable contributions has also been observed [23] in a recent analysis of *B*- decays using the FAT approach

We have also observed that present experimental data *B* -decays clearly show the presence of FSI strong phase differences, which agrees with other analysis carried out for these decays [20, 21]. We also wish to mention that we are aware of a work by Sharma and Katoch [36] who have assumed ratio of the non-factorizable amplitudes equal to -0.828 in the absence of experimental data at that time, and predicted sum of branching fractions of $\bar{B}^0$ -decays, which is not in agreement with the latest experimental measurements. Moreover, their approach is different from ours, as they have not considered the final state interaction effects.